\documentclass[12pt,preprint,a4paper]{aastex}
\usepackage{epstopdf}
\usepackage{graphics,epsf}
\usepackage{amsmath}                
\usepackage{amsfonts}               
\usepackage{amssymb}                
\usepackage{epsfig}                 

\def \cm{~\rm{cm}}

\def \s{~\rm{s}}
\def \km{~\rm{km}}

\def \K{~\rm{K}}
\def \g{~\rm{g}}

\def \AU{~\rm{AU}}
\def \erg{~\rm{erg}}

\def \yr{~\rm{yr}}

\def \pc{~\rm{pc}}

\def \dyn{~\rm{dyn}}

\def \mol{~\rm{mol}}

\title{Using Intermediate-Luminosity Optical Transients (ILOTs) to reveal extended extra-solar Kuiper belt objects}

\author{Ealeal Bear\altaffilmark{1} and Noam Soker\altaffilmark{1}}

\altaffiltext{1}{Department of Physics, Technion -- Israel Institute of
Technology, Haifa 32000 Israel; ealeal@physics.technion.ac.il, soker@physics.technion.ac.il}

\begin{document}

\begin{abstract}
We suggest that in the rare case of an Intermediate-Luminosity Optical Transient (ILOTs) event, evaporation of extra-solar Kuiper belt objects (ExtraKBOs) at distances of $d \approx 500 - 10000\AU$ from the ILOT can be detected. If the ILOT lasts for $1$ month to a few years, enough dust might be ejected from the ExtraKBOs for the IR emission to be detected. Because of the large distance of the ExtraKBOs, tens of years will pass before the ILOT wind disperses the dust. We suggest that after an ILOT outburst there is a period of months to several years during which IR excess emission might hint at the existence of a Kuiper belt analog (ExtraK-Belt).
\end{abstract}


\section{INTRODUCTION}
\label{sec:intro}
 The Kuiper belt in our solar system contains icy bodies (Kuiper belt objects - KBOs) from Neptune to about $50\AU$ (e.g., \citealt{Petitetal1999}). These icy bodies are in fact cometary nuclei. Short period comets have orbital parameters that suggest that their formation is from the Kuiper belt (e.g., \citealt{Melnicketal2001, Sternetal1990}). The total mass of the belt in the annulus $30<a<50 \AU$ is estimated to be in the range of $0.1 - 0.26 M_\oplus$ (e.g., \citealt{Jewittetal1998,Chen2006}). Around $10^5$ KBOs posses a diameter larger than $100\km$, the biggest being Pluto with $D_{\rm Pluto}\approx 2400\km$. There are more objects with smaller diameter than $100\km$  (e.g., \citealt{Melnicketal2001}).
The small icy bodies in the solar system motivate observations and theoretical studies of such bodies orbiting other stars (e.g., \citealt{Wyatt2008} for a review). For that, we first note the rich variety of the properties of the Kuiper belt objects in the solar system.

The Kuiper belt of the solar system contains different sub groups of icy bodies, such as the cold population (inclination of less than 4 degrees to the ecliptic plane) and the hot population (inclination of more than 4 degrees to the ecliptic plane). The hot population was formed closer to the Sun and was transported into the Kuiper belt. The cold population formation is still under debate. The two main scenarios are local formation and transportation to its current position (e.g., \citealt{Gomes2003, LevisonMorbidelli2003, Levisonetal2008, Morbidellietal2014} and earlier references therein).
 The belt is usually divided into an inner belt ($a<40\AU$) and an outer belt ($a>41\AU$); the area of $40<a<42\AU$ is dynamically unstable due to secular resonances (e.g., \citealt{Petitetal1999}).
As noted by \cite{BrownPan2004} the plane of the Kuiper belt is affected by many parameters such as the total angular momentum of the solar system, recent stellar encounters, and unseen distant masses in the outer solar system.
 Therefore, it is quite possible that some stars have very extended Kuiper-like belt. For example, a massive, $\approx 10 M_\odot$, disk extending to $\approx 1000 \AU$ was detected recently around the young O-type Star AFGL~4176  \citep{Johnstonetal2015}.
 The rich variety of icy bodies in the solar systems and the different phases of late stellar evolution raise the possibility to detect icy bodies, or their outcome, during some particular stages of stellar evolution.

Belts similar to the Kuiper belt exist in extra-solar planetary systems (e.g., \citealt{Wyatt2008}), like \rm{HR}$~8799$ (e.g., \citealt{SadakaneNishida1986, Suetal2009, Matthewsetal2014, Controetal2015}). They are detected through dusty debris resulting from collisions \citep{Greavesetal2005}. The first discovery was due to an excess emission towards Vega, an A0V star \citep{Aumannetal1984} which spectral energy distribution corresponds to cold dust grains at $80\AU$ from the star \citep{Wyattetal2004}. \cite{Wyattetal2004} studied stars (and binary systems) within $6\pc$ from the Sun and found that $\epsilon$~\rm{Eri} and  $\tau$~\rm{Cet} have derbris disks of  $0.01M_\odot$ at $60\AU$ and $ 5\times 10^{-4} M_\odot$ at $55\AU$, respectively. Other detections, include $\zeta^2$~\rm{Ret} \citep{Eiroaetal2010} and the G stars HD 38858 and HD 20794 \citep{Kennedyetal2015}.

The subject of KBOs analogs (ExtraKBOs) around post-main-sequence (MS) stars has been investigated before by \cite{Jura2004}. They discuss the ice sublimation of ExtraKBOs when the star reaches the red giant branch (RGB) phase. This rapid sublimation leads to a detectable infrared excess at $25 \mu m$, depending on the mass of the ExtraKBOs.

In the present study we investigate the evaporation of ExtraKBOs during a transient brightening event called intermediate-luminosity optical transients (ILOTs). ILOTs are eruptive outbursts with total kinetic energy of $10^{46} - 10^{49}\erg$ that last weeks to several years (\citealt{KashiSoker2015} and references there in). Their peak bolometric luminosity can be of the order of $10^{40} \erg\s^{-1}$ (e.g., ILOT NGC 300OT had $L_{\rm bol} = 1.6 \times 10^{40} \erg \s^{-1}$, \citealt{Bondetal2009}).

ILOTs that harbor AGB or extreme-AGB (ExAGB) pre-outburst stars, were modeled for a single (e.g., \citealt{Thompsonetal2009}) and binary systems (e.g., \citealt{KashiSoker2010, SokerKashi2011,SokerKashi2012}).
 The progenitor masses of the AGB ILOTs, M85 OT2006-1 and NGC 300 OT2008-1 were estimates as  $M_1 < 7 M_\odot$ and $6 < M_1 < 10 M_\odot$, respectively \citep{Ofeketal2008, Prietoetal2009}. \cite{McleySoker2014} conclude that such ILOTs are most likely powered by a binary interaction. More generally, ILOTs are thought to be powered by gravitational energy released in a high-accretion rate event in binary stars \citep{KashiSoker2015}, even possibly during a grazing envelope evolution \citep{Soker2016}. The required orbital separation is about 1-3 times the giant radius, at least during the periastron passages. So the general binary separation at periastron passages (which is the orbital radius for circular orbits) for these systems is in the general range of $0.5 - 5 \AU$. Relevant to the binary model is the suggestion made by \cite{Humphreysetal2011}  that the ILOT NGC~300~2008OT-1 had a bipolar outflow; see however the claim by \cite{Adamsetal2015} that NGC~300~2008OT-1 might have been a SN explosion rather than an ILOT event.

ILOTs powered by merging stars can occur in MS stars, such as V838~Mon \citep{SokerTylenda2003, TylendaSoker2006}. Since the luminosity, mass-loss rate and dust production in winds from MS stars are several orders of magnitudes below those of AGB stars, the IR excess due to the ExtraKBO evaporation will be much more prominent. This holds true despite ILOT events of MS stars being generally shorter than those expected from AGB stars. The merger product itself becomes very red and form dust, but this dust is hotter and concentrated in the center (e.g., \citealt{Munarietal2002, BondSiegel2006, KaminskiTylenda2013, Chesneauetal2014, Kaminskietal2015}).

In the present study we estimate the IR excess due to sublimation of ExtraKBOs near an ILOT event. The sublimated gas from the ExtraKBOs carries dust which might be detected, we claim, throughout an IR excess. Such observations will reveal the properties of Kuiper belt analogs (ExtraK-Belt) around systems that experience ILOT events (that might hint at a binary merger).
We start our study by calculating the sublimation of ExtraKBOs when the star experiences an ILOT event (Sec. \ref{sec:ExtraKBOS_sub}), and then in section \ref{sec:IR} we address the emission and possible ways to detect this type of evaporation.
A short summary of our main conclusions is given in section \ref{sec:summary}.

\section{ILOT SUBLIMATION OF EXTRA-SOLAR KUIPER BELTS (ExtraK-Belt)}
\label{sec:ExtraKBOS_sub}
\subsection{Sublimation rate}
\label{subsec:sub}

ExtraKBOs/comets have not been directly observed around post-main sequence (MS) stars. Observation around subgiants have been made, and although no ExtraKBOs have been found, the large amount of dust indicate that dusty material can survive the MS phase (for more details, see \citealt{Bonsoretal2013, Bonsoretal2014}).
\cite{Sternetal1990} and \cite{SaavikFordNeufeld2001} model the ice sublimated from comets with orbital separation of more than $100\AU$ from stars that evolved along the AGB. \cite{Sternetal1990} study the detectability of comet clouds during the post-main-sequence phase of stellar evolution and show that the change in luminosity in the post MS stage has a dramatic effect on the reservoirs of comets.  They calculate the temperature (see, \citealt{Sternetal1990}, eq. 2) at different orbital separations from the post MS star and find out that KBOs will go through intensive mass-loss of water and more volatile species. Objects in the Oort cloud will sublimate volatile species, and depending on the eccentricity of their orbit might go through intensive water sublimation as well. Comets on circular orbits will sublimate mostly volatile species such as ${\rm CO_2}$, ${\rm CO}$, ${\rm NH_3}$, while comets on eccentric orbits that come closer to the star will go through a significant water sublimation. Their model takes into account a star which evolves from the MS to the red giant phase over $\approx 10^8\yr$. At the RGB phase its luminosity is $\approx 300 L_\odot$, a change that increases the water sublimation radius from $2.5\AU$ at $L=L_\odot$ to $\approx 40\AU$ at $L\approx 300L_\odot$.

Once the hydrogen has exhausted and the core contracted, a core-helium flash starts the horizontal branch (HB) phase (the helium flash is inside the core and is too short to influence sublimation). The AGB phase that follows the HB phase lasts for $\approx 10^7\yr$ with a luminosity of several thousands solar luminosity. At this stage the water sublimation radius increases to the Kuiper belt and the inner Oort cloud as well. \cite{Melnicketal2001} agree that the AGB stage affects the Kuiper belt but argue that the Oort cloud is not affected from an increase in luminosity in the range of $100 - 3000 L_\odot$ when the star evolves along the AGB.

\cite{SaavikFordNeufeld2001}, that studied \rm{IRC}$~+10216$, and \cite{Maerckeretal2008}, who studied  M-type AGB stars, argued that the presence of water vapor in carbon rich AGB stars possibly suggests the existence of extra-solar cometary systems.

There is a direct connection between the sublimation rate and the chance of ExtraKBOs detection \citep{Jura2004}.
We calculate the sublimation rate per unit area per object from the kinetic theory model as customary (e.g., \citealt{DelsemmeMiller1971} eq. 7; \citealt{CowanAhearn1979} eq. 2; \citealt{FanaleSalvail1984} eq. 8; \citealt{PrialnikBar-Nun1990} eq. 17;  \citealt{vanLieshoutetal2014} eq. 12;  references therein)
\begin{equation}\label{eq:z}
\dot z =P_{\rm vap}(T_{\rm ExtraKBO}) \sqrt{\frac{m}{2\pi R_g T_{\rm ExtraKBO}}} \g \s^{-1} \cm^{-2},
\end{equation}
where $m$ is the molecular weight in $\g \mol^{-1}$, $R_g$ is the gas constant, $P_{\rm vap}(T_{\rm ExtraKBO})$ is the vapor pressure in  $\dyn \cm^{-2}$ and $T_{\rm ExtraKBO}$ is the ExtraKBO`s temperature. Note that the sublimation rate per object is a function of the ExtraKBO area. We define the sublimation rate for a single ExtraKBO
 \begin{equation}\label{eq: Z_S}
  \dot Z_S = \dot z \pi R_{\rm ExtraKBO}^2,
 \end{equation}
where $\pi R_{\rm ExtraKBO}^2$ is the area of the ExtraKBO facing the ILOT. Asteroids and comets are subject to a size distribution, e.g., as the one used by \cite{Jura2004}, but the mass is probably concentrated in the most massive bodies. For asteroids disruption near white dwarf (WD) it is customary to take radii in the range of $R_{\rm ast}\approx 1-1000\km$ (e.g., \citealt{Jura2003, Farihietal2014, Verasetal2014}). In this paper we scale expression with $R_{\rm ExtraKBO}=5\km$.
In order to calculate the sublimation rate per object, we must know the vapor pressure which is derived from the temperature of the ExtraKBO and its composition. We address below three molecules, ${\rm H_2O}$, ${\rm CO}$ and ${\rm CO_2}$ and assume that the ExtraKBO is composed entirely of one type of molecule.
\subsection{Vapor pressure}
\label{subsec:pvap}
       The vapor pressure for volatile species such as ${\rm CO}$ (e.g., \citealt{Bujarrabaletal2013, Hillenetal2015}) and ${\rm H_2O}$ can be described by the $Clausius~–~Clapeyron$ relation for sublimation (e.g., \citealt{Prialnik2006, RosenbergPrialnik2009})
    \begin{equation}\label{eq:Cl-Cl}
      \frac{d \ln P_{\rm vap}}{d T_{\rm ExtraKBO}}=\frac{H}{R_g T_{\rm ExtraKBO}},
    \end{equation}
    where $H$ is  the latent heat (enthalpy) of sublimation (calculated from table 1). We integrate eq. \ref{eq:Cl-Cl} and derive (e.g., \citealt{Gombosietal1985, LichteneggerKomle1991, Prialnik2006, RosenbergPrialnik2009, Gronkowski2009b})
  \begin{equation}\label{eq:int-cl}
  P_{\rm vap} = A\exp(-B/T_{\rm ExtraKBO}),
\end{equation}
    where the relevant constants for three different molecules are given in Table 1 and are derived from CRC Handbook of Chemistry and Physics (where tables of $P_{\rm vap}$ and $T$ are given for each molecule; \citealt{Lide2015}).
    \newline
    Table 1
     \begin{tabular}{|l|c|c|}
       \hline
        Molecule & $A [\dyn \cm^{-2}]$             & $B [\K]$ \\
        \hline
        ${\rm CO}$   & $1.75 \times 10^{11}$ &  946.91   \\
         \hline
        ${\rm H_2O}$ & $3.69 \times 10^{13}$ &  6151.7 \\
         \hline
        ${\rm CO_2}$ & $1.14 \times 10^{13}$ &  3157.7   \\
                 \hline
     \end{tabular} \label{table:Gas_pvap}
           \newline
            \newline
\subsection{ExtraKBO temperature}
\label{subsec:T_extraKBo}

The temperature of ExtraKBOs can be calculated from the energy
equation, similar to that used for comets (e.g.,
\citealt{Beeretal2006, Gronkowski2009a, PrialnikRosenberg2009}):
        \begin{equation}\label{eq:energy}
         E_{\rm ILOT}=E_{\rm thermal} + E_{\rm sub} + E_{\rm con},
        \end{equation}
     where:
        \begin{equation}\label{eq:E_ILOT}
          E_{\rm ILOT} = \frac{(1-A_{\rm ExtraKBO})L_{\rm ILOT}}{4\pi d^2} ,
        \end{equation}
is the heating flux from the ILOT outburst that is absorbed by the
body, $A_{\rm ExtraKBO}$ is the ExtraKBO albedo, $L_{\rm ILOT}$ is the
ILOT luminosity and d is the orbital separation. The terms on the
right hand side of equation (\ref{eq:energy}) are as follows.
       \begin{equation}\label{eq:E_thermal}
          E_{\rm thermal}=\epsilon \sigma T_{\rm ExtraKBO}^4,
        \end{equation}
is the cooling by thermal radiation per unit area of the object
where, $\epsilon$ is the emissivity and $\sigma$ is the Stefan
Bolzman constant,
       \begin{equation}\label{E_sub}
         E_{\rm sub} = H \dot z,
       \end{equation}
   is cooling per unit area by sublimation. Due to the very low
temperatures the energy conductivity term $E_{\rm con}$ is
insignificant for ${\rm H_2O}$, ${\rm CO}$ and ${\rm CO_2}$, and
will be neglected here. We note that the area of evaporation and cooling can be $4\pi R_{\rm ExtraKBO}^2$  depending if the heat had time to spread over the ExtraKBO. Although rotation is possible, since the conductivity is negligible, we assume that most of the cooling and sublimation is from the surface facing the ILOT. Therefore, we take the effective area for energy and mass-loss of one object to be
 $\pi R_{\rm ExtraKBO}^2$.

 The supervolatile {\rm CO}, that is known
to exist in comets, can be in its gaseous form even in the low
temperatures of the Kuiper belt. The {\rm CO} sublimation rate for
several known KBOs (such as 2060 Chiron,1998 WH24, 7066 Nessus and
others) are calculated to be in the order of $\dot Z_s \approx
10^{28} \rm{molecules} \s^{-1}$ (for more details, see table 6 in
\citealt{BockeleeMorvanetal2001}). Due to ${\rm CO}$ low
absorption band it has only been detected on Pluto and Triton
(e.g., \citealt{SchallerBrown2007, Brown2012} and references
within). Sublimation of ${\rm CO}$ can be significant for ExtraKBOs
around ILOTs (Fig. \ref{fig:T_Z_M_d}
below).

The dust temperature is calculated by balancing radiation heating
and cooling. When the temperature exceeds the sublimation
temperature, the dust is destroyed. The sublimation distance of
dust particles from an energy source is given by (e.g.,
\citealt{NetzerLaor1993, LaorDraine1993})
\begin{equation}\label{eq:Rsub}
  R_{\rm sub-dust} \approx 24 \left( \frac{L_{\rm ILOT}}{10^6 L_\odot} \right)^{\frac{1}{2}}\AU. 
\end{equation}
As we are interested in ExtraK-Belt at distance of $\gg 100 \AU$, the
dust that is carried out by the sublimating gas from the ExtraKBOs survives and does not sublimate.

\subsection{ExtraKBO Sublimation by ILOTs}
\label{subsec:sub ILOT}

\begin{figure}
  \vspace{-4cm}
\includegraphics[scale=0.7]{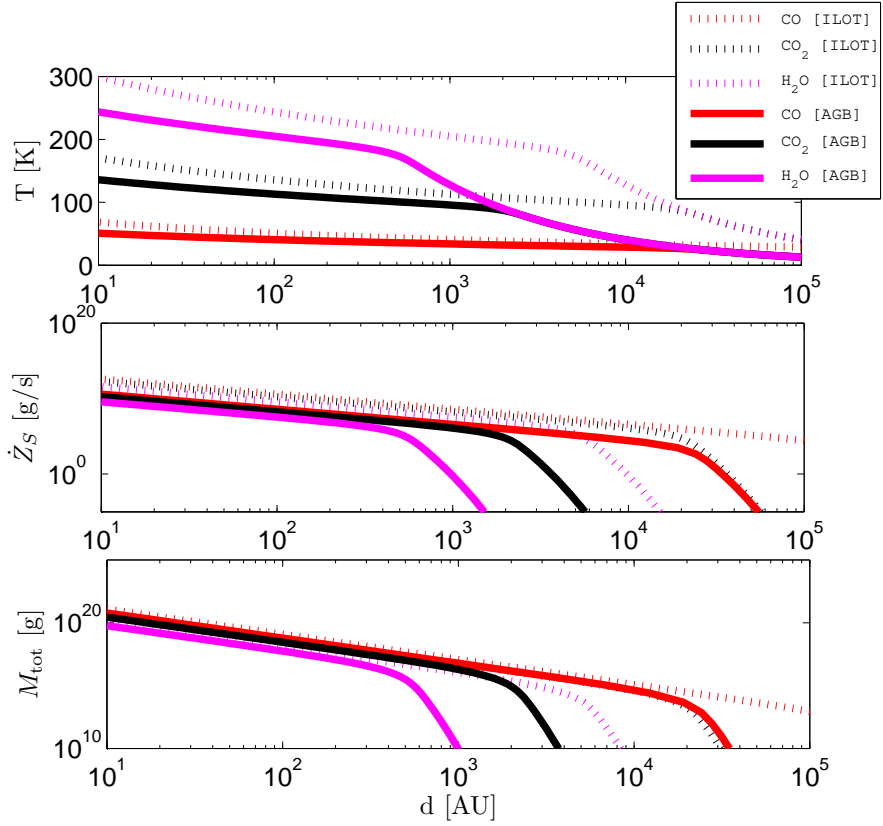}
   \vspace{6cm}
\caption{Properties of an ExtraKBO during an ILOT event and during the AGB phase, as function of its the distance from the ILOT. The upper panel represents ExtraKBO temperature,
the middle panel depicts ExtraKBO sublimation rate per object for $R_{\rm
ExtraKBO}=5\km$ and the lower panel represents total mass
ejected from this object. Two options are considered and
represented by different lines: ExtraKBO around ILOT (dotted lines)
and ExtraKBO around AGB stars (solid lines). Red ,black and pink
represent ExtraKBO made from single molecules ${\rm CO}$,${\rm
CO_2}$ and ${\rm H_2O}$ respectively. The vapor pressure
parameters for all three molecules for all panels are taken from
table 1. For the lower panel, we take the duration of the ILOT
event and the AGB effective duration as $t_{\rm ILOT}=10\yr$ and $t_{\rm AGB}=500\yr$,
respectively.}
 \label{fig:T_Z_M_d}
 \end{figure}

In order to simplify the calculation we assume that the ExtraKBO is
composed of one type of gas only. In Fig. \ref{fig:T_Z_M_d} we present the
ExtraKBO temperature as a function of orbital separation from the
ILOT according to equation (\ref{eq:energy}). For the ILOT
event we scale quantities with a luminosity of $L_{\rm ILOT}= 10^6 L_\odot$  and
a duration of $t_{\rm ILOT}=10\yr$; later we will substitute $t_{\rm ILOT}=0.1 \yr$ for the case of ILOTs of MS stars.
For the upper AGB phase of the
star, we take $L_{\rm AGB}= 10^4 L_\odot$  and an effective duration of
$t_{\rm AGBw}=500 \yr$.  The effective duration of the AGB star is the time span of the wind in the ExtraK-Belt location
   \begin{equation}
  \label{eq:tAGBw}
   t_{\rm AGBw} \approx 500  \left(\frac {d}{1000 \AU}\right)
    \left( \frac {v_w}{10 \km \s^{-1}}   \right)^{-1}  \yr,
   \end{equation}
where $v_w$ is the AGB wind speed (e.g., \citealt{Wintersetal2003}). AGB dust that was lost earlier will be further away and be cooler. Since we can not differentiate the dust carried by the AGB wind from the dust carried by the gas sublimating from the ExtraKBOs once they are at the same region, we take this effective AGB duration.

\begin{figure}
  \vspace{-4cm}
\includegraphics[scale=0.7]{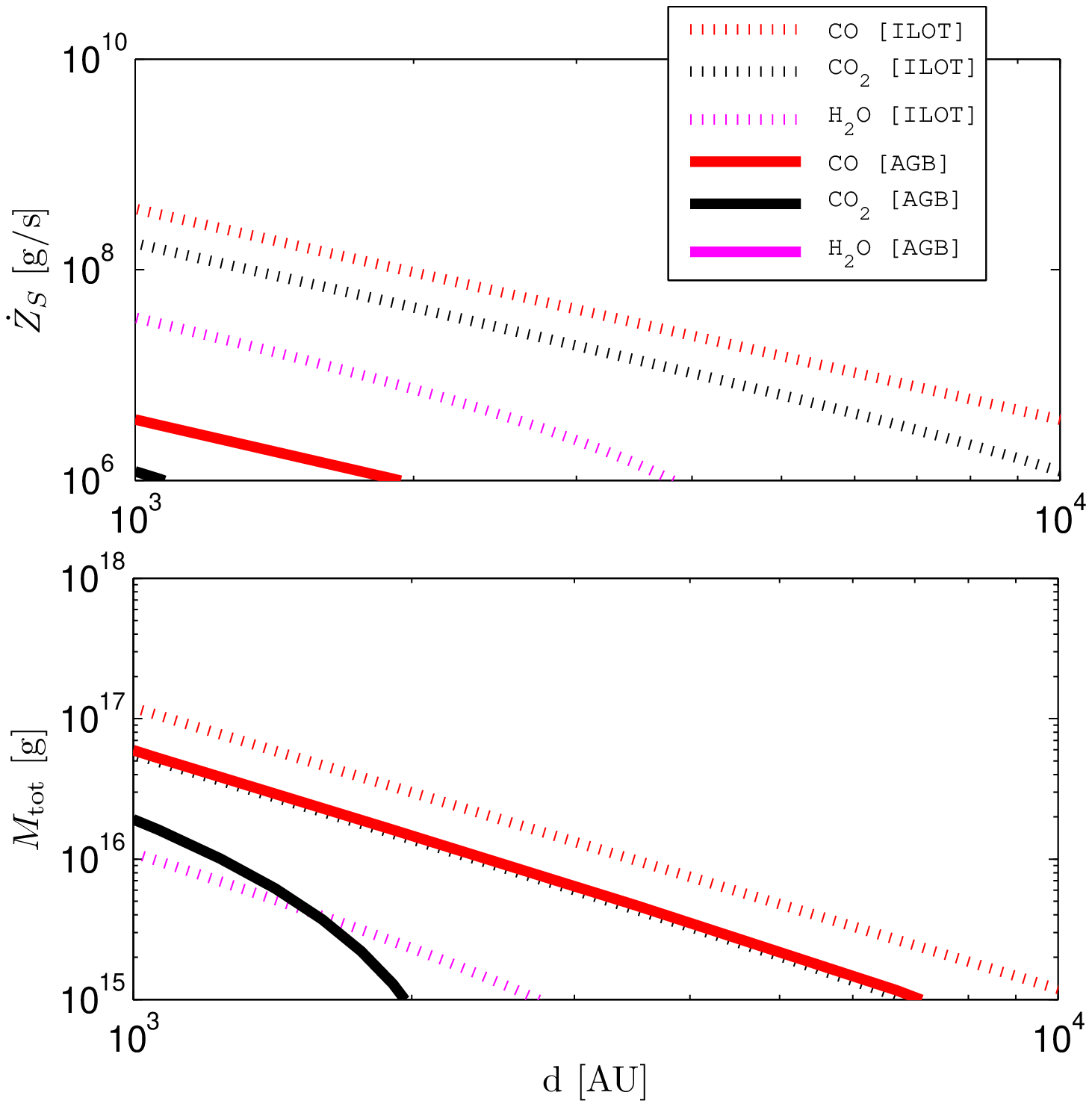}
   \vspace{6cm}
\caption{The same as the corresponding panels in Fig \ref{fig:T_Z_M_d}, but zooming on the distances range of $10^3-10^4\AU$. }
 \label{fig:zoom}
 \end{figure}

We compare the effects of the AGB star and the ILOT event on the
temperature (Fig. \ref{fig:T_Z_M_d}, upper panel), evaporation rate per object with a radius of $R_{\rm ExtraKBO}=5 \km$ (Fig. \ref{fig:T_Z_M_d}, middle panel) and
total mass evaporated from the object (Fig. \ref{fig:T_Z_M_d}, lower panel). In the lower panel we calculate the evaporated mass during the AGB effective duration given by equation (\ref{eq:tAGBw}) and during the ILOT outburst ($10 \yr$). The luminosities are given above,  $L_{\rm AGB}= 10^4 L_\odot$ and $L_{\rm ILOT}= 10^6 L_\odot$, respectively. In Fig. \ref{fig:zoom} we zoom on the distance range $d = 10^3 -10^4\AU$.

Figure \ref{fig:T_Z_M_d} indicates that the temperature of a single ExtraKBO around the
ILOT event is, as excepted, higher than the temperature for the AGB phase
for all compositions. For the sublimation rate per object we can see that for each molecule there is "knee" (gap) where the sublimation rate for ExtraKBO around the AGB star substantially drops while for the ILOT event it is still significant. For ExtraKBOs residing at these distances a prominent IR excess is expected due to ILOT sublimation. We emphasis that this IR excess is not a result of the dust from the ILOT wind but a result of the dust carried by the sublimating gas from the ExtraKBOs. This gap in mass-loss is  observed in the lower panel  which represents the total gas mass from a single ExtraKBO. The interesting distance ranges of the ExtraKBO where the influence of the ILOT is much larger than that of the AGB star are as follows $d \approx 10^3-10^4\AU$ for ${\rm H_2O}$, $d \approx 3 \times 10^3-2 \times 10^4\AU$ for ${\rm CO}$
and $d \ga 3\times 10^4\AU$ for ${\rm CO_2}$ , as can be seen from Figs. \ref{fig:T_Z_M_d} and \ref{fig:zoom}. Since the ExtraKBOs are
not composed from a single gas it is hard to predict the exact interesting distance ranges. However, observing the distance of $\approx 1000 -
10000\AU$ after an ILOT event during the first few months to years might reveal excess amount of dust and might hint on the presence of an ExtraK-Belt. For the purpose of our calculations we choose to look specifically at a distance of $d=1000\AU$.

We scale the ExtraK-Belt total mass $M_{\rm Belt}$ with the
solar system one $0.1 M_\oplus$, (e.g., \citealt{Gladmanetal2001}), so that
the number of objects in the ExtraK-Belt is
\begin{equation}
  \label{eq:Nobj}
  N_{\rm obj}\approx 10^{9}
  \left(\frac{M_{\rm Belt}}{0.1M_\oplus}\right)
  \left(\frac{\rho_{\rm ExtraKBO}}{1 \g \cm^{-3}}\right)
  \left(\frac{R_{\rm ExtraKBO}}{5 \km}\right)^{-3},
\end{equation}
where $\rho_{\rm ExtraKBO}$ is the average density of the ExtraKBOs.
We find that the total mass of the sublimated dust and gas from
ExtraK-Belt in the outburst around an ILOT event are
\begin{equation}\label{eq:M_dust_tot}
M_{\rm dust-ILOT} = \eta  M_{\rm gas-ILOT}
  \approx 5\times 10^{-9}   \eta
 \left(\frac{\dot{z_S}_{(1000\AU)}}{3.5 \times 10^{7} \g \s^{-1}}\right)
 \left(\frac{t_{\rm ILOT}}{10\yr}\right)
 \left(\frac{\gamma}{1}\right)
 \left(\frac{N_{\rm obj}}{10^9}\right)
 M_\odot,  
 \end{equation}
where $\eta$ is the dust to gas ratio, that is estimated from comets to be $\eta \approx
0.1 - 10$  (e.g., \citealt{Fulleetal2010, Laraetal2011,
Schmidtetal2015}), and $\gamma$ is a scaling factor depending on whether the ILOT event occurs during the AGB or the MS phase. For a MS star the duration of the ILOT event is only $\approx 0.1 \yr$, so that $\gamma=0.01$.
The distance of the ExtraK-Belt was taken at $d=1000\AU$ and the sublimation rate per object ($\dot{Z_S}$) is calculated for ${\rm H_2O}$. The ILOT luminosity was taken to be $L_{\rm ILOT} = 10^6 L_\odot$.
The average total mass-loss rate from the ExtraK-Belt is $\dot
M_{\rm gas-ILOT} = \dot Z_S  N_{\rm obj}$.

The ratio of the dust mass from the ILOT to the AGB star is an important quantity
given by
     \begin{equation}
     \begin{split}
     \label{d_WIND_ILOT_AGB}
       \frac{M_{\rm dust-ILOT}}{M_{\rm AGBw}} =
      N_{\rm obj}\left(\frac{\dot Z_s}{ \dot M_{\rm AGBw}}\right)
       \left(\frac{t_{\rm ILOT}}{t_{\rm AGBw}}\right)
        \left(\frac{\eta_{\rm ILOT}}{\eta_{\rm AGB}}\right) \\
       \approx
        10^{-3} 
 \left(\frac{\dot{z_S}_{(1000\AU)}}{3.5 \times 10^{7} \g \s^{-1}}\right)
  \left(\frac{\dot M_{\rm AGBw}}{10^{-6} M_\odot \yr^{-1}}\right)^{-1}
 \left(\frac{N_{\rm obj}}{10^9}\right)
  \left(\frac{t_{\rm ILOT}}{10\yr}\right)\\
   \left(\frac{t_{\rm AGBw}}{500 \yr}\right)^{-1}
    \left(\frac{\eta_{\rm ILOT}}{1}\right)
      \left(\frac{\eta_{\rm AGB}}{0.01}\right)^{-1}
,
  \end{split}
     \end{equation}
     where $\dot z_s$ is the sublimation rate of the ILOT at $1000\AU$ (for $\rm{H_2O}$), $\dot M_{\rm AGBw}$ is the mass-loss rate of the wind and $\eta_{\rm ILOT}$ and $\eta_{\rm AGB}$ are the dust to gas ratio of the ILOT and AGB respectively.
   Despite the much lower sublimated dust from the ILOT than that residing in the AGB wind, for the parameters chosen here, the sublimated dust from the ExtraK-Belt is concentrated in a particular radius, and hence will have strong emission at a particular temperature. It might possibly be detected, but we estimate that only if the AGB wind has a lower mass-loss rate than chosen here.

More promising might be ILOTs from MS stars (or slightly evolved off the MS), such as V838~Mon, V1309~Sco, and similar events (e.g., \citealt{Tylendaetal2011, Kaminskietal2015, Pejchaetal2016}). In MS stars the wind mass-loss rate is much lower and dust production is less efficient than in AGB stars. Taking the ratio of dust mass for an ILOT and a MS star gives
\begin{equation}
 \begin{split}
     \label{d_WIND_ILOT_MS}
       \frac{M_{\rm dust-ILOT}}{M_{\rm MSw}} \approx
       5000
 \left(\frac{\dot{z_S}_{(1000\AU)}}{3.5 \times 10^{7} \g \s^{-1}}\right)
  \left(\frac{\dot M_{\rm MSw}}{10^{-12} M_\odot \yr^{-1}}\right)^{-1}
 \left(\frac{N_{\rm obj}}{10^9}\right)\\
  \left(\frac{t_{\rm ILOT}}  {0.1 \yr } \right)
    \left(\frac{t_{\rm MS}}  {10 \yr } \right)^{-1}
   \left(\frac{\eta_{\rm ILOT}}{1}\right)
      \left(\frac{\eta_{\rm MS}}{0.001}\right)^{-1},
       \end{split}
     \end{equation}
where $\dot M_{\rm MSw}$ is the wind mass-loss rate from the MS star, $\eta_{\rm MS}$ is the gas to mass ratio of the MS star, and $t_{\rm MS}$ is the effective duration of the mass-loss phase. Because of the faster MS wind compared to the AGB wind, $v_{\rm MSw} \approx 500\km \s^{-1}$, and the closer ExtraK-Belt, about $1000 \AU$, the effective duration is shorter (eq. \ref{eq:tAGBw}).

\section{IR excess}
\label{sec:IR}

Our solar system contains the asteroid belt and the Kuiper belt.
Some of the dust observed in our solar system results from collisions in the asteroid belt which form 'hot' dust (zodical) at $2-4\AU$ at $\approx 270\K$ and cold dust at $30-50\AU$ from the Sun and at a temperature of $50-60\K$ which is a result of KBOs collision.
Observation of ExtraK-Belts is not simple. \cite{Beichmanetal2006} note that inferred IR excess for the solar system Kuiper belt is
$L_{\rm IR}/L_\odot \approx 10^{-7}- 10^{-6}$ while for the asteroid belt is $L_{\rm IR}/L_\odot \approx  10^{-7}$ (e.g., \citealt{Beichmanetal2006} and references within).

The IR excess due to the dust released from gas sublimation can be crudely estimated assuming a uniform distribution of $1 \mu m$ - size dust grains \citep{Jura2004}
\begin{equation}\label{eq:IR_excess}
\frac{ L_{\rm excess}}{L_{\rm ILOT}} \approx \frac{\chi M_{\rm dust-ILOT}}{4 \pi d^2},
\end{equation}
where the opacity is estimated as \citep{Jura2004}
\begin{equation}\label{eq:chi}
  \chi= \frac{\pi a_{\rm dust}^2}{4 \pi \rho_{\rm dust} a_{\rm dust}^3/3}\approx 7500\left(\frac{\rho_{\rm dust}}{1 \g \cm^{-3}}\right)^{-1}\left(\frac{a_{\rm dust}}{1 \mu m}\right)^{-1} \cm^2 \g^{-1},
\end{equation}
where $a_{\rm dust}$ is the dust radius.
We note that \cite{Jura2004} takes into account different models for the mass-loss rate of KBOs. He differentiates between large and small populations of KBOs and follow their evolution along with the change in luminosity of the host star. Since our outburst is short and we assume that its duration is only $0.1-10\yr$, we do not differentiate and simply take the sublimation rate per object from the energy equation (eq. \ref{eq:z}) for a uniform distribution of ExtraKBOs. Substituting  eq. (\ref{eq:chi}) into eq. (\ref{eq:IR_excess}) yields
\begin{equation}\label{eq: IR_excess_frac}
\frac{ L_{\rm excess}}{L_{\rm ILOT}} \approx 10^{-5} \left(\frac{M_{\rm dust-ILOT}}{5 \times 10^{-9}M_\odot}\right) \left(\frac{d}{1000 \AU}\right)^{-2}\left(\frac{a_{\rm dust}}{1 \mu m}\right)^{-1}\left(\frac{\rho_{\rm dust}}{1 \g \cm^{-3}}\right)^{-1}.
\end{equation}
It serves as a lower limit and can be higher depending on composition, grain size, ILOT duration and mass of the ExtraK-Belt.
\cite{RiviereMarichalaretal2014} observe a tiny IR excess
of $L_{\rm IR}/L_\odot \approx 2 \times 10^{-6}$ around the young star $\rm{HD}~ 29391$ (a F0IV star, \citealt{AbtMorrell1995}), that is part of the beta Pictoris moving group (BPMG). This IR excess suggests a derbis disk with a lower limit on the inner radius of $82\AU$.
We note that \cite{Eiroaetal2010} report an initial result of the presence of a dust ring at $\approx 100\AU$ from the solar type star $\zeta^2$~\rm{Ret} with $L_{\rm excess}/L_{\rm *} \approx 10^{-5}$.

\cite{Wyatt2008} estimates the fractional luminosity from a collision cascade (orbital eccentricities greater than $10^{-3}$ to $10^{-2}$) in his eq. (15). Substituting ExtraKBOs with a diameter of $\approx 10 \km$ and total mass of $M_{\rm tot}=0.1M_\oplus$ as in our eq. \ref{eq:Nobj}, we find the fractional IR luminosity to be $f_{\rm coll} \approx 10^{-8} (r/1000 \AU)^{-2}$.  This is three orders of magnitude lower than the IR emission produced from the sublimated ExtraKBOs (our eq. \ref{eq: IR_excess_frac}). Furthermore, calculating the timescale for mutual collisions,  using the parameters given in \cite{Wyatt2008} for a fitted population of debris disks around A stars, and our belt mass of $M_{\rm tot}=0.1M_\oplus$, we find that the timescale to establish collision cascade is $\approx 3 \times 10^9 \yr$. This implies that ExtraK-Belt around stars with an initial mass larger than about $1.5 M_\odot$ will not have time to establish the cascade, and the values of $f_{\rm coll}$ will be lower even than $10^{-8}$.
Therefore, IR emission from a collision cascade is insignificant relative to IR excess due to sublimation of ExtraKBOs.  We note that a few single collisions can occur, but they  are rare and result in an insignificant IR excess (see, eq. 19 in \citealt{Wyatt2008}).  

 The observations of an evaporated ExtraK-Belt at hundreds of AU and more from an ILOT event will be quite difficult. The low temperature of the dust implies that most of the radiation will be at longer wavelength than the Spitzer band. Detecting rotational emission lines from CO molecules is not expected to be sensitive enough. \cite{Matraetal2015} used ALMA to observe the debris disk of Fomalhaut for CO lines, and set an upper limit on the CO mass there. For cases studied here the emission is expected to be even lower, as both the evaporated mass is lower than their upper limit, and there is no H$_2$ molecules to collisionally excite CO. Hence luminosity in CO will be very weak.
It seems that the most optimistic situation will be for a future mission comparable or better than the Wide-field Infrared Survey Explorer (WISE). If an ILOT event takes place during the operation of such a far IR space observatory, then it might be a prime target for a search of an evaporated ExtraKBO.

\section{SUMMARY}
\label{sec:summary}
In the present study we studied the influence of ILOT events, in both post-main sequence (MS) stars and MS stars, on very small substellar objects. We considered the sublimation of comets and similar objects, Kuiper belt analogues (ExtraK-Belts), by rare cases of ILOTs that might last weeks to years. Such ILOTs can take place in AGB stars, (e.g., NGC 300OT \citealt{Monard2008, Bergeretal2009, Bondetal2009}), and during earlier phases of the evolution, including the main-sequence.
Although ILOT events are rare, metal pollution of WDs show that many planetary systems survive the post-MS evolution ({e.g., \citealt{Jura2006, Juraetal2007, Jura2008, Debesetal2012, Farihietal2011, Farihietal2012, Melisetal2012, Farihietal2014}). So, we expect that sublimation of comet-like bodies as far as few thousands AU from the star will take place in ILOT events, and dust will be released (eq. \ref{eq:M_dust_tot}).

If the ILOT event lasts for several weeks (in MS stars) to several months and years (in AGB stars), enough dust might be ejected from these bodies (eqs. \ref{d_WIND_ILOT_AGB} and \ref{d_WIND_ILOT_MS}) for the IR emission to be detected (eq. \ref{eq: IR_excess_frac}). This dust is concentrated at the ExtraK-Belt and therefore might be distinguished from the dust formed in the stellar wind. Although a star spends millions of years on the AGB with high luminosity, there is a region where ILOT events will release gas and dust from comets, but not AGB stars (Fig. \ref{fig:zoom}).

The extra dust that might be observed at thousands of AU from the ILOT event is the dust formed by the sublimating ExtraKBOs. Lots of dust is formed in the ILOT event, but it will take it tens of years or longer to reach the ExtraK-Belt region discussed here. The excess IR luminosity should be searched for months to years after the ILOT event. The phenomenon can be observed where the influence of the ILOT is much larger than that of the AGB star. In section \ref{sec:IR} we noted that the release of dust by collisions will be small and will not contribute significantly to the IR excess. Although ILOTs themselves turn red, their typical temperature at early times is hotter than that of the sublimated dust at large distances studied here. So spectroscopically the sublimated dust can be distinguished from the IR emission of the ILOT event itself.
At distances of thousands of AU from the ILOT, such IR emission might be even resolved with future instrumentation.

ILOT events in MS stars last only several weeks. However, the MS stellar mass-loss rate and dust production in the wind are several orders of magnitudes below those of AGB stars. Therefore, the IR excess due to the ExtraKBOs evaporation might be much more prominent than for ILOTs in AGB stars.
 Due to the clean-of-dust region around many MS stars (relative to the region around AGB stars), in particular for stars in the halo of the Galaxy, ILOT events of MS stars in the distance past might lead to small IR excess today. The evaporated dust might survive for a relatively long time, long after the star has returned to the MS after the ILOT event. Due to uncertainties, we do not quantify the expected IR excess in this case.

To summarize, we encourage the search for evaporated Kuiper-like belts around ILOTs within several years after outburst.

We thank an anonymous referee for helpful comments.
This research was supported by the Asher Fund for Space Research at the Technion, and the E. and J. Bishop Research Fund at the Technion.

\end{document}